\documentclass[preprint,12pt]{elsarticle}

\usepackage{amssymb}

\usepackage{multirow}

\journal{Annals of Nuclear Energy}

\begin{document}

\begin{frontmatter}



\title{Modeling of critical experiments and its impact on integral covariance matrices and correlation coefficients}


\author{Elisabeth Peters, Fabian Sommer, Maik Stuke}
\ead{elisabeth.peters@grs.de, fabian.sommer@grs.de, maik.stuke@grs.de}
\address{Gesellschaft f\"ur Anlagen- und Reaktorsicherheit (GRS) gGmbH, Forschungszentrum, Bolzmannstr. 14, 85748 Garching (M\"unchen), Germany}

\begin{abstract}
In this manuscript we study the modeling of experimental data and its impact on the resulting integral experimental covariance and correlation matrices. By investigating a set of three low enriched and water moderated UO$_2$ fuel rod arrays we found that modeling the same set of data with different, yet reasonable assumptions concerning the fuel rod composition and its geometric properties leads to significantly different covariance matrices or correlation coefficients. Following a Monte Carlo sampling approach, we show for nine different modeling assumptions the corresponding correlation coefficients and sensitivity profiles for each pair of the effective neutron multiplication factor $k_{\rm eff}$. Within the 95$\%$ confidence interval the correlation coefficients vary from 0 to 1, depending on the modeling assumptions. Our findings show that the choice of modeling can have a huge impact on integral experimental covariance matrices. When the latter are used in a validation procedure to derive a bias, this procedure can be affected by the choice of modeling assumptions, too.  The correct  consideration of correlated data seems to be inevitable if the experimental data in a validation procedure is limited or one cannot rely on a sufficient number of uncorrelated data sets, e.g. from different laboratories using different setups etc.           
\end{abstract}

\begin{keyword}
Criticality Safety \sep Code Validation \sep Correlation Coefficients \sep Covariance Matrices \sep Monte Carlo Sampling 


\end{keyword}

\end{frontmatter}


\section{Introduction}
\label{Intro}

Criticality safety assessments require a prediction of the effective neutron multiplication factor (k$_{\rm eff}$) below a sufficient safety margin. This predicted value is derived using a validated calculation method with validated computer codes, e.g. so called criticality codes to calculate the k$_{\rm eff}$ of an application case. The validation of a criticality code can be achieved by recalculations of suitable critical experiments performed in laboratories and documented and evaluated e.g. in \cite{ICSBEP}. In recent years, several authors discussed the fact that depending on the application case and the choice of experiments, the effect of correlated experimental data on the determination of the bias, its uncertainty, and the resulting safety margins has to be considered \cite{Ivanova,NCSDCorr, NCSDBias, Hoefer2015, ICNCCorr, ICNCBias, Sobes, BJICNC, MONK}. The questions arising in the field of determination and handling of integral experimental covariance matrices in the process of code validation are also discussed in the Expert Group on Uncertainty Analysis for Criticality Safety Assessment (UACSA), a sub-group of the Working Party on Nuclear Criticality Safety (WPNCS) of the Nuclear Energy Agency (NEA) within the Organization for Economic Co-operation and Development (OECD). Actual questions which arose recently are: How to treat given sets of similar experimental data without knowing all exact statistical dependencies; and further, what are the implications on modeling these experiments in a code validation procedure regarding the consideration of the complete integral experimental correlation or covariance matrices? 

In this manuscript we address these questions by following parts of the group’s proposal for a benchmark called Role of Integral Experiment Covariance Data for Criticality Safety Validation \cite{UACSABMIV}. In contrast to the benchmark proposal we focus on a reduced number of experiments but a total of nine different modeling approaches. 

With the following analysis we add a new perspective to the discussion and show the effect of different modeling approaches for the same set of experimental data on the resulting integral covariance or correlation matrices.

Correlated data can arise if different experiments share parts of the experimental setup, measurement systems, or other relevant parameters. Some experiments described in the ICSBEP are not performed as single experiments, but slight variations of a setup were repeatedly investigated and published as a series of the same experiment. This is e.g. the case for LEU-COMP-THERM- 039 (LCT-39), where the number and location of empty positions  in a fuel rod grid were varied. In the following work we focus on the experimental data from experiments numbers 6, 7, and 8 from this series described in detail in \cite{ICSBEP, UACSABMIV, ICNCCorr} and references therein. The critical experiments consist of water moderated low enriched uranium fuel rods with a thermal neutron spectrum. The experimental setups are 22$\times$22 arrays consisting of 363 (459, 448) fuel rods for experiment 6 (7, 8) and 121 (25, 36) empty spots, respectively. For further details we refer to \cite{ICSBEP, ICNCCorr}. Clearly these experiments share certain components, and treating them as individual statistical independent data sets in the process of validation probably would not be appropriate. Hence, the determination of the integral covariance or correlation matrix of the experiments is a crucial step on the way to determine a bias of the calculated application case k$_{\rm eff}$. 
%
%
%
%
\section{Methods and Parameters}
\label{Methods}

For the determination of the integral covariance matrices of k$_{\rm eff}$ and the corresponding correlation matrices we use a Monte Carlo Sampling approach, and SUnCISTT \cite{SUnCISTT} to steer and evaluate the numerous SCALE 6.1.2 \cite{SCALE} calculations. For two sets A and B of n sampled neutron multiplication factors k$_{\rm eff}$, the covariance ${\rm cov}^{\rm AB}$ is defined as 
\begin{equation}
{\rm cov}^{\rm AB}=\frac{1}{n-1}\sum_{i=1}^n{ \left({k_{\rm eff}}^{{\rm A},i}-\overline{{k_{\rm eff}}}^{{\rm A},i}\right)
 \left({k_{\rm eff}}^{{\rm B},i}-\overline{{k_{\rm eff}}}^{{\rm B},i}\right) }
\label{Cov}
\end{equation}
with $\overline{{k_{\rm eff}}}$ symbolizing the expectation value of $k_{\rm eff}$, in our case the sample mean. The covariance can be interpreted as a measure of how much the $k_{\rm eff}$ of the two sets change simultaneously. A positive covariance indicates the following  monotonic connection between the two sets: large (or low) values in A correspond to large (or low) values in B. A negative covariance indicates the opposite behavior: large values in A correspond to low values in B. Due to its linearity the covariance gives only a tendency of the connection of two sets of random variables. To get comparable statements for more than two sets the covariance can be normalized with the standard deviation $\sigma$ to get the correlation coefficient ${\rm cor}$:
\begin{equation}
{{\rm cor}}^{\rm AB}=\frac{1}{\sigma_A \sigma_B}{\rm cov^{{\rm AB}}}
\label{ck}
\end{equation}
The correlation coefficient is a dimensionless measure of the linear dependence of two sets of random variables and takes values between +1 (complete positive linear connection) and -1 (complete negative linear connection). The confidence interval around the calculated ${\rm cor}$ is due to the value limitation of ${\rm cor}$ non-symmetric and must be determined by using transformations to so called Fisher’s distribution $z$ \cite{Fisher}. The latter is almost normally distributed and depends on the sample correlation coefficient:
\begin{equation}
z({{\rm cor}})=\frac{1}{2} {\rm ln}\left(\frac{1+{\rm cor}}{1-{\rm cor}}\right)
\label{zck}
\end{equation}
The corresponding tolerance intervals are calculated via $z\pm{\rm CL}\times\delta$ with the confidence level CL (e.g. 1.96 for the 95$\%$ confidence interval) and the sample standard deviation $\delta=\sqrt{(n-3)^{-1}}$.
The resulting z values are then transformed back to ${\rm cor}$ values. All given confidence intervals in the following work are 95$\%$ intervals.

Following a Monte Carlo Sampling approach, each value describing the experiment has to be interpreted as a distribution function. This means in turn, that the definition and interpretation of the experimental parameters and their uncertainties is essential. It strongly depends on the quality of the experimental data and availability of precise uncertainty specifications. To circumvent the problem of determining suitable distribution function for each parameter, we apply the ones given in \cite{UACSABMIV} which are also listed in \ref{Tab.1}. All experimental parameters are supposed to follow a uniform ${\rm U}$(a,b) or normal distribution N($\mu$,$\sigma$). Assuming the three experiments LCT-39 6, 7, and 8 to be statistical independent gives a correlation coefficient close to zero. Results for this assumption are shown for the correlation of k$_{\rm eff}$ values calculated by KENO V.a using the parameters given in table \ref{Tab.2} for 250 Monte Carlo samples for each experiment. The underlying model assumptions for the results of Figure 1 are very simple and straight forward: It is assumed, that the fuel rods are all identical in composition and position within its unit cell. In consequence, the modeling of one experiment consists basically of a $22\times22$ array of identical unit cells for the fuel rods and the empty positions respectively. 
\begin{figure}[ht]
	\centering
		\includegraphics[width=.75\textwidth]{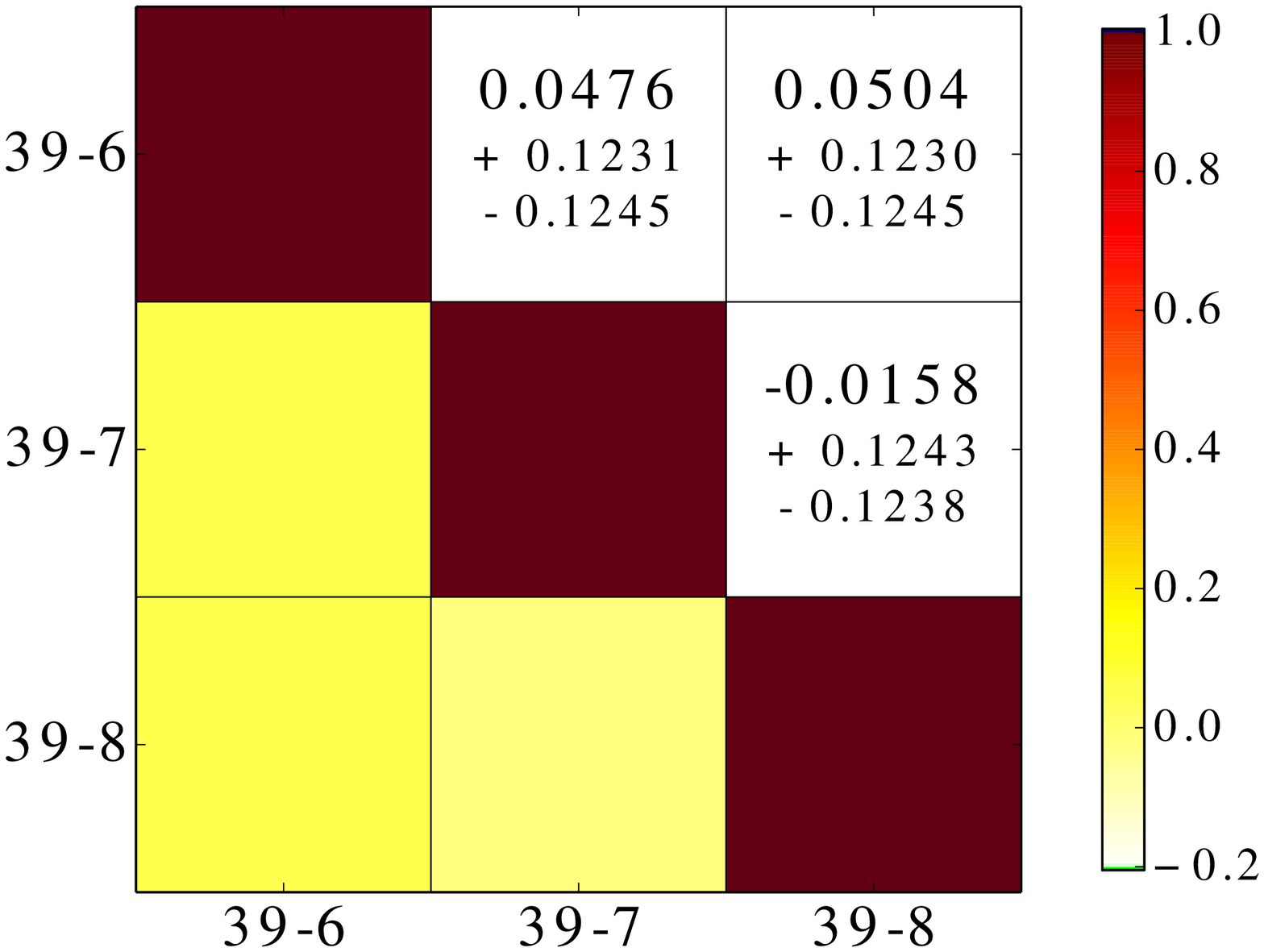}
	\caption{ Correlation coefficients of the experiments LCT-39 6, 7, and 8 assuming all fuel rods to be identical and no statistical dependence between each experimental setup.}
	\label{fig:Pic1}
\end{figure}

\begin{table}[hp]
	\centering
		\begin{tabular*}{.99\textwidth}{@{\extracolsep{\fill}}|p{5.2cm}|p{2cm}|r|}
			\hline
{\bf Model parameters} 	& {\bf Type of variation} 		&	{\bf Distribution functions} \\ \hline\hline
			Fuel diameter [cm]		 	& depends on scenario	& N(0.7892, 0.0017) \\ \hline
			Fuel lengths [cm]			&	depends on scenario	& N(89.7, 0.3) \\ \hline
			Fuel density [g/cm$^3$]			& depends on scenario	& N(10.38, 0.0133 ) \\ \hline
			Fuel content 234U	[At.-$\%$] &depends on scenario	& N(0.0307, 0.0005 )  \\ \hline
			Fuel content 235U [At.-$\%$]	&depends on scenario	& N(4.79525, 0.002)  \\ \hline
			Fuel content 236U	[At.-$\%$]&depends on scenario	& N(0.1373, 0.0005)  \\ \hline
			Boron concentration [atom/barn$\times$cm$\times10^{-8}$]&	depends on scenario	& N($6.9037$ , 
$0.8)$ \\ \hline
			Critical water height [cm] &	individual	& N($\mu,\sigma$) dep. on experiment\\ \hline
			Angle of fuel rod	& individual	&U( 0, 2$\pi$) \\ \hline
			Offset of grid hole x [cm]	&individual	& N(0, 0.00742) \\ \hline
			Offset of grid hole y	[cm] &individual	& N(0, 0.00742)  \\ \hline
			Hole diameter	[cm] &depends on scenario	& N(0.0105, 0.0085)  \\ \hline
			Inner cladding diameter	 [cm] &depends on scenario	&U (0.81, 0.83)\\ \hline
			Cladding thickness [cm] &	depends on scenario	&U (0.055, 0.065)    \\ \hline
		\end{tabular*}
	\caption{All model parameters and their distribution characteristics, following the suggestions of the benchmark proposal \cite{UACSABMIV}.}
	\label{Tab.1}
\end{table}

\begin{table}[htp]
	\centering
		\begin{tabular*}{.99\textwidth}{@{\extracolsep{\fill}}|l|r|p{4cm}|}
			\hline
			{\bf Code} & {\bf Parameter} & {\bf Value}\\ \hline
			{\multirow{4}{*} {KENO V.a}} & Nuclear data library & ENDF/B-VII (ce)	\\ 
															& Neutrons per generation & 10,000 \\
															& Skipped generations & 500 \\
															& $\sigma_{\rm MC}$ & 5$\times10^{-4}$(Sc. A to D); 1$\times10^{-4}$(Sc. E to H)\\ \hline
			SUnCISTT   & Number of samples & 250 \\ \hline
	
		\end{tabular*}
	\caption{Used codes and cornerstones of calculations. KENO V.a is taken from the CSAS5 sequence of SCALE 6.1.2}
	\label{Tab.2}
\end{table}
%
\section{Modeling Assumptions}
\label{Modeling}

Having determined all relevant parameters and their distribution functions, a calculation model is built to calculate the neutron transport equations and determine the neutron multiplication factor. Obviously the model should be as close as possible to the experimental setup to get reasonable results. However, in the statistical interpretation of experimental series like the one investigated in this article, the available data might leave some freedom of choice. The results shown in figure \ref{fig:Pic1} represent a model simplification of the experimental setup by assuming all fuel rods in one sample to be identical. However, it appears to be more reasonable that due to manufacturing tolerances of the experimental equipment individual fuel rods may vary in both, their individual composition and position within the unit cell. The position of the fuel rod in the unit cell is then limited by the grid hole. For some simplicity we assume the fuel rod to be always vertical, meaning a 90 degrees angle to the horizontal plane. 
\begin{figure}
	\centering
		\includegraphics{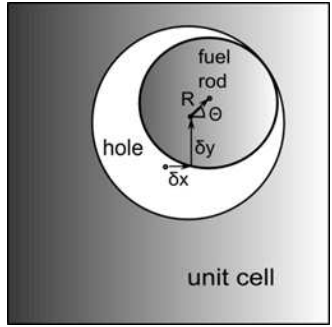}
	\caption{Modeling approach for the displacement of grid hole and fuel rod in a unit cell (not to scale)}
	\label{fig:FR_Model}
\end{figure}
The modeling approach for the fuel rod displacement is depicted in figure \ref{fig:FR_Model}. For the modeling of the experiments in KENO V-a, this implies each fuel rod to be simulated within its own unit cell, which we assume to have fixed dimension for all fuel rods. According to figure \ref{fig:FR_Model}, position of the grid hole might be displaced from the center of the unit cell by $\delta_x$   and $\delta_y$ in x and y direction. The center of the fuel rod itself might again be displaced in x- and y-direction, denoted by the radial displacement  R and angle $\theta$. In our modeling approach R is indirectly defined by the assumption that the fuel rod is in contact with the grid hole. 

\begin{figure}[hp]
	\centering
		\includegraphics[width=0.75\textwidth]{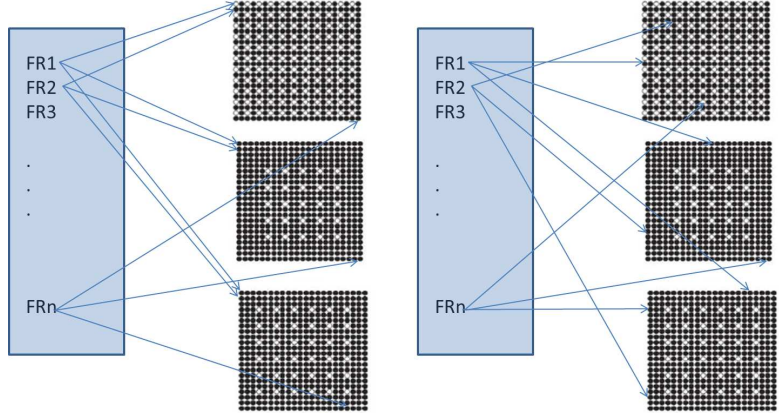}
	\caption{Sketch of modeling approaches for scenarios F (left) , G and H (right). The blue boxes represent the fixed library of individual generated fuel rods. The black and white squares represent the three experimental setups LCT-39 6, 7, and 8 from top to bottom. A black dot represents a fuel rod, a white one an empty spot. The simulation corresponding to the left part of the picture assumes a fixed position for each fuel rod in each experimental setup. E.g. FR1 is always in the top left spot for every experiment. The right part of the figure depicts the assumption of each fuel rod being randomly placed in the grid for each experiment.}
	\label{fig:Scetch_ScF+G}
\end{figure}

We chose eight different modeling approaches, scenarios A to H, depending on assumptions on the fuel similarity and position of each single fuel rod. Scenarios A to E assume the fuel to be identical for all fuel rods in all experiments. One can argue that this might be a reasonable approximation, based on the assumption of a very accurate fuel fabrication process with only tiny tolerances. However, scenarios F and G assume a set of the maximal needed fuel rods (484 for the 22$\times$22 grid array), each statistical independent. These fuel rods are placed for all experiments in a fixed position for scenario F or randomly for each experiment in scenario G, see figure \ref{fig:Scetch_ScF+G}. Finally, scenario H assumes all fuel rods in every experiment to be statistical independent. The statistical dependence of the fuel between two experiments decreases from scenario E to H. The results shown in figure \ref{fig:Pic1} assume the same modeling assumptions as scenario A but any correlations were neglected, meaning no statistical dependence between each experimental setup. This scenario is named ‘NoCor’.
\begin{table}[hp]
	\centering
		\begin{tabular*}{.99\textwidth}{@{\extracolsep{\fill}}|l|p{2cm}|p{2cm}|p{1.8cm}|p{1.8cm}|p{1.8cm}|}
			\hline
			{ Scenario} & { Grid hole displacement } &{ Grid hole diameter } &{ 
			Inner cladding diameter} &{ Cladding thickness } &{ Fuel variation } \\ \hline \hline
			NoCor & no; rod centered & shared  & shared & shared  & shared \\ \hline
			A			&	no; rod centered  & shared  & shared & shared  & shared  \\ \hline 
			B			& $\delta_x$,$\delta_y$ & shared & shared	& shared &	shared \\ \hline
			C			&$\delta_x$,$\delta_y$	& individual &	shared &	shared &	shared \\ \hline
			D			&$\delta_x$,$\delta_y$  &	individual &	individual	&	shared	& shared \\ \hline
			E			&$\delta_x$,$\delta_y$  &	individual & 	individual	&	individual 	&	shared \\ \hline
			F			&$\delta_x$,$\delta_y$	&	individual	&	individual	&	individual	&	484 FR, fixed pos. \\ \hline
			G			&$\delta_x$,$\delta_y$	&	individual	&	individual	&	individual	&	484 FR, random pos.\\ \hline
			H			&$\delta_x$,$\delta_y$	&	individual	& individual	&	individual	&	individual\\ \hline
		\end{tabular*}
	\caption{Modeling assumptions for fuel rod geometries and compositions of one sample for each different scenario. The variation of fuel in the last column means the variation of the diameter, length, density and enrichment of the fuel as well as the boron impurity. Scenario NoCor (scenario H) is identical to A (scenario G), except for neglecting statistical dependencies between experiments.}
	\label{Tab.3}
\end{table}

%
%
\section{Results}
\label{Results}

The analysis allover required a total of 6.750 SCALE inputs with up to 20.000 lines per input file. The calculations were performed using a total of 55.000 CPU-h and 882 TByte-h. The results then were processed and statistically analyzed using SUnCISTT. 

\begin{figure}[hp]
	\centering
		\includegraphics[width=0.75\textwidth]{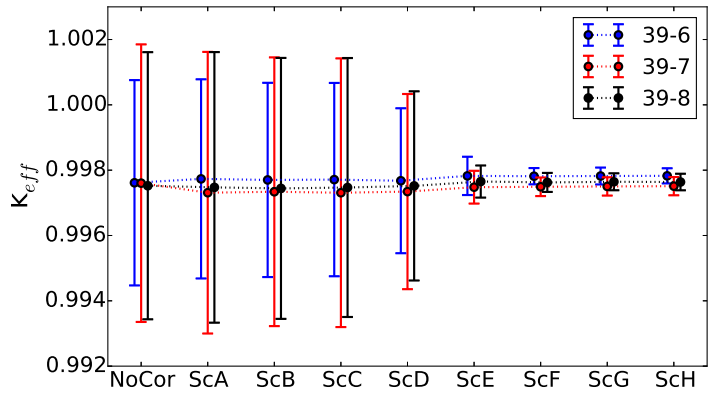}
	\caption{Resulting $k_{\rm eff}$ values for 250 Monte Carlo samples for each experiment and scenario. The error bars indicate the standard deviation. While the nominal values remain fairly constant for all scenarios, the standard deviation decreases significantly from scenario D to F.}
	\label{fig:keffs_StdDev}
\end{figure}

The resulting $k_{\rm eff}$ values for each experimental data set and modeling scenario are shown in figure \ref{fig:keffs_StdDev}. We found a good agreement within the 2-$\sigma$ range of the experimental data ${k_{\rm eff}}^{\rm exp}=1.0$ ($\pm0.0012$ for exp. 7,8) ($\pm0.0009$ for exp. 6) given in \cite{ICSBEP}) and our results. The SCALE calculations with the applied continuous energy library ce\_v7\_endf (based on ENDF/B-VII) in the CSAS5 sequence systematically underestimates $k_{\rm eff}$, which is a known effect for low enriched uranium setups \cite{SCALEVal}. The larger error bars of the Monte Carlo approach of Scenario A in comparison to the error propagation approach done in the ICSBEP Handbook are not attributed to a general difference between the two methods. They rather arise from a different interpretation of the system parameter uncertainties. In the original experiment description \cite{Valduc} the uncertainty of the inner cladding diameter ($\pm$ 0.01 cm) and the cladding thickness ($\pm$ 0.005 cm) are reported to be independent. The uncertainty of the outer cladding diameter is obtained by error propagation. In the ICSBEP evaluation, the uncertainty of the cladding thickness is split equally between inner and outer diameter. This results in an uncertainty of the outer cladding diameter of $\pm$ 0.0025 cm, which reduces its impact on the uncertainty of $k_{\rm eff}$ significantly. The original evaluation assumes further a Gaussian distribution by dividing the half tolerance by $\sqrt{3}$. The resulting distributions for the outer cladding diameter and their impact on $k_{\rm eff}$ are different for both considerations. However, the values based on the original literature were used and for further details we refer to \cite{ICNCCorr}. Scenarios E to H in which the individual variations of the parameters partly cancel out each other, have significant lower error bars.

\begin{figure}[hp]
	\centering
		\includegraphics[width=0.95\textwidth]{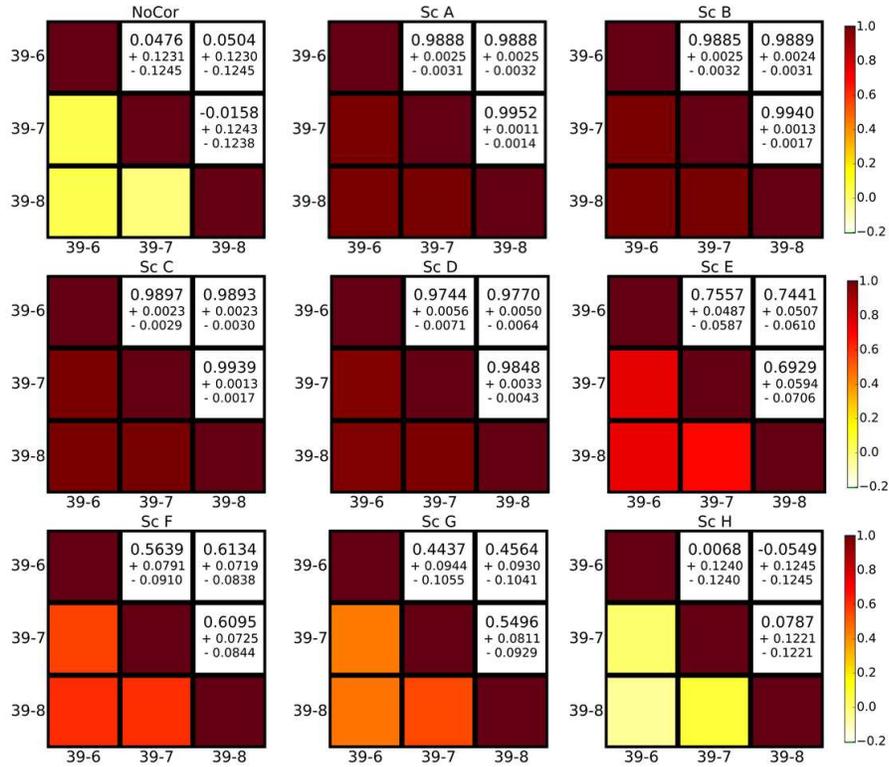}
	\caption{Matrices of the correlation coefficients for the 9 scenarios. The upper part of each matrix shows the ck values and the ranges for the 95$\%$ confidence interval. The lower part shows the nominal ${\rm cor}$ values color coded. The scale varies from dark red for ${\rm cor}$  = 1 to white for ${\rm cor}$ = -0.2. The resulting correlation coefficients for scenario NoCor and H are 0 within the 95$\%$ confidence interval. Scenarios A to D show ${\rm cor}$ values close to 1, while scenarios E to G show a slight decreasing of ${\rm cor}$ from approximately 0.75 to 0.5.}
	\label{fig:All_9_Sc_CI}
\end{figure}

In figure \ref{fig:All_9_Sc_CI} nine colored plots are shown for the ${\rm cor}$ values of the experiments LCT-39 6, 7, and 8 as well as the ${\rm cor}$ value and the 95$\%$ confidence interval.

The results show correlation coefficients around 0 for the scenarios NoCor and H, as expected, since there are no relevant parameters with shared values between the individual experiments. Note that the difference between the NoCor and H scenario is the variation of the fuel rods: In contrast to NoCor, in scenario H each fuel rod in each experiment is simulated individually and statistically independent. This difference is mapped in the sensitivity plots in figure \ref{fig:Fig7} and \ref{fig:Fig8} which show the correlation coefficients of each parameter with the resulting $k_{\rm eff}$. Note, that the performed sensitivity analysis shows the impact of the actual variation of each parameter on the $k_{\rm eff}$ uncertainty. However, we do not perform a sensitivity analysis by varying only one parameter at a time. This means, our sensitivities depend on the chosen distribution functions and their characteristics. Changing these assumptions in our approach might lead to a different sensitivity profile. This approach was chosen since we are interested in determining the contribution of each varied input parameter on the uncertainty of $k_{\rm eff}$ for given modeling assumptions. 

While for the scenario NoCor the most relevant parameters are the cladding inner radius and thickness, the only important parameter for scenario H is the critical water height. It is notable, that in this case the different interpretations of the given experimental data lead to comparable ${\rm cor}$ values but totally different sensitivity profiles.  The highest correlation coefficients for scenario NoCor are the ones for the cladding inner diameter and thickness, and for the radius of the fuel. The only dominant parameter for scenario H is the critical water height.
Scenarios A, B, C and D show all ${\rm cor}$ values close to 1 with only little deviations between the different ${\rm cor}$ values of the scenarios. The ${\rm cor}$ values for scenario A to C are even the same within the 95$\%$ confidence interval. Their corresponding sensitivity profiles show huge similarities: The three largest ${\rm cor}$ values are the cladding inner radius and thickness and the radius of the fuel. For scenario B and C the U-235 weight-$\%$ plays a more prominent role. The sensitivity profiles for scenario D show a different behavior since the inner cladding diameter here is varied individually for each fuel rod. The leading contribution to the sensitivity profile now solely results from the cladding thickness.

Scenario E shows correlation coefficients of approximately 0.75 between the experiments 6 and 7 and 0.7 between experiments 7 and 8. The difference between the ${\rm cor}$ values is due to lower number of fuel rods in experiment 6 compared to 7 and 8. Thus, the individual variation of the cladding inner radius and thickness for each fuel rod affects the correlation coefficient of the experiments 7 and 8 more. The corresponding sensitivity profile shows the fuel radius as the leading parameter. A mild impact is shown by the fuel density and critical water height (additional the weight-$\%$ for U-235 and the fuel height for LCT-39 7).

Scenarios F and G show significantly smaller correlation coefficients between the experiments. The difference to Scenario E is that now also the fuel content of each fuel rod is varied individually. This can be seen in the sensitivity profile of both scenarios in figure \ref{fig:Fig8}, where the dominant parameter is the critical water height.

\begin{figure}[htbp]
	\centering
		\includegraphics[width=0.75\textwidth]{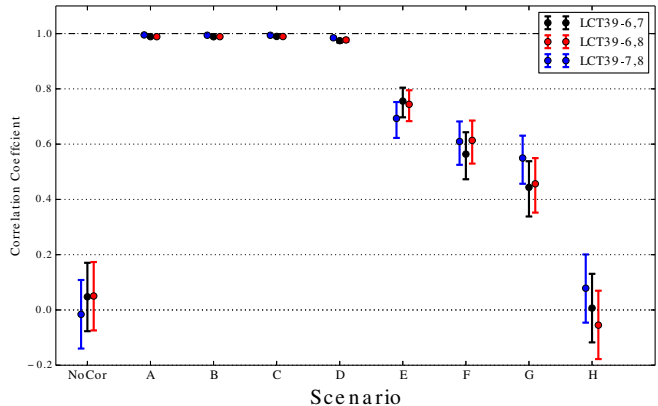}
	\caption{Correlation coefficients for the pairs of experiment 6 and 7 (black), 6 and 8 (red), and 7 and 8 (blue) of Series LCT-39 for the 9 scenarios. The error bars represent the 95$\%$ confidence interval. Scenarios A, B, C, and D show high correlation coefficients, all close to 1. For scenarios E to G the coefficients decrease to approximately 0.7, 0.6, and 0.5 respectively. The correlation coefficients for scenario H are 0 within the 95$\%$ confidence interval, like the ones for the NoCor scenario.}
	\label{fig:ScVsCk}
\end{figure}
\begin{table}[hp]
	\centering
	\begin{tabular*}{.99\textwidth}{@{\extracolsep{\fill}}|r|l|r|l|}
			\hline
					Key word 	&	Parameter	&	Key word	&	Parameter \\ \hline\hline
				rad\texttt{\_}CladIn	&	cladding inner raduis	&	w$\%$\texttt{\_}U235	& weight-$\%$ U-235 \\ \hline
				{thick\texttt{\_}Clad}	&	cladding thickness	&		w$\%$\texttt{\_}U236	&	weight-$\%$ U-236 \\ \hline
				{rad\texttt{\_}Fuel}		&	fuel radius	&	{height\texttt{\_}Water}	&	water height \\ \hline
				{height\texttt{\_}Fuel}	&	fuel height	&	{rad\texttt{\_}Hole}	&	hole radius \\ \hline
				{dens\texttt{\_}Fuel} 	&	fuel density	&	{delta\texttt{\_}Hole\_X}	& $\delta_x$ \\ \hline
				{dens\texttt{\_}B10}		&	B-10 density	&	{delta\texttt{\_}Hole\_Y}	&	$\delta_y$ \\ \hline
				w$\%$\texttt{\_}U234	&	weight-$\%$ U-234	&	{angle\texttt{\_}Rod}	&	$\theta$ \\ \hline

		\end{tabular*}
	\caption{Parameters and key words used in figure \ref{fig:Fig7} and \ref{fig:Fig8}}
	\label{Tab.4}
\end{table}

\begin{figure}[hp]
	\centering
		\includegraphics[width=0.95\textwidth]{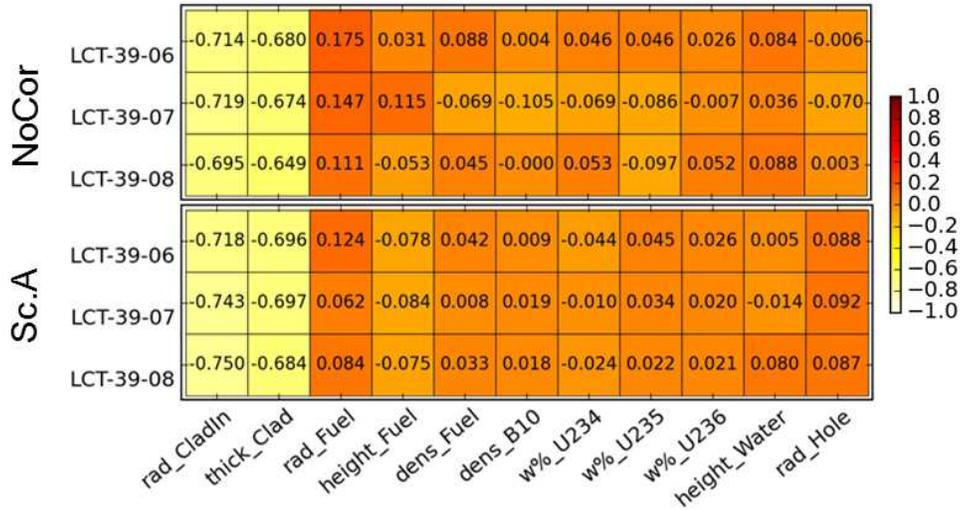}
	\caption{ Correlation coefficients of each individual model parameter listed in table 4 and the resulting 250 $k_{\rm eff}$ values for scenarios NoCor and A. The profiles are similar due to identical modeling assumptions. The only difference between the two scenarios is the assumption of correlations in scenario A. Note the different scale of the color coded representation of ${\rm cor}$ w.r.t. figure \ref{fig:All_9_Sc_CI}.}
	\label{fig:Fig7}
\end{figure}

\begin{figure}[hp]
	\centering
		\includegraphics[width=0.95\textwidth]{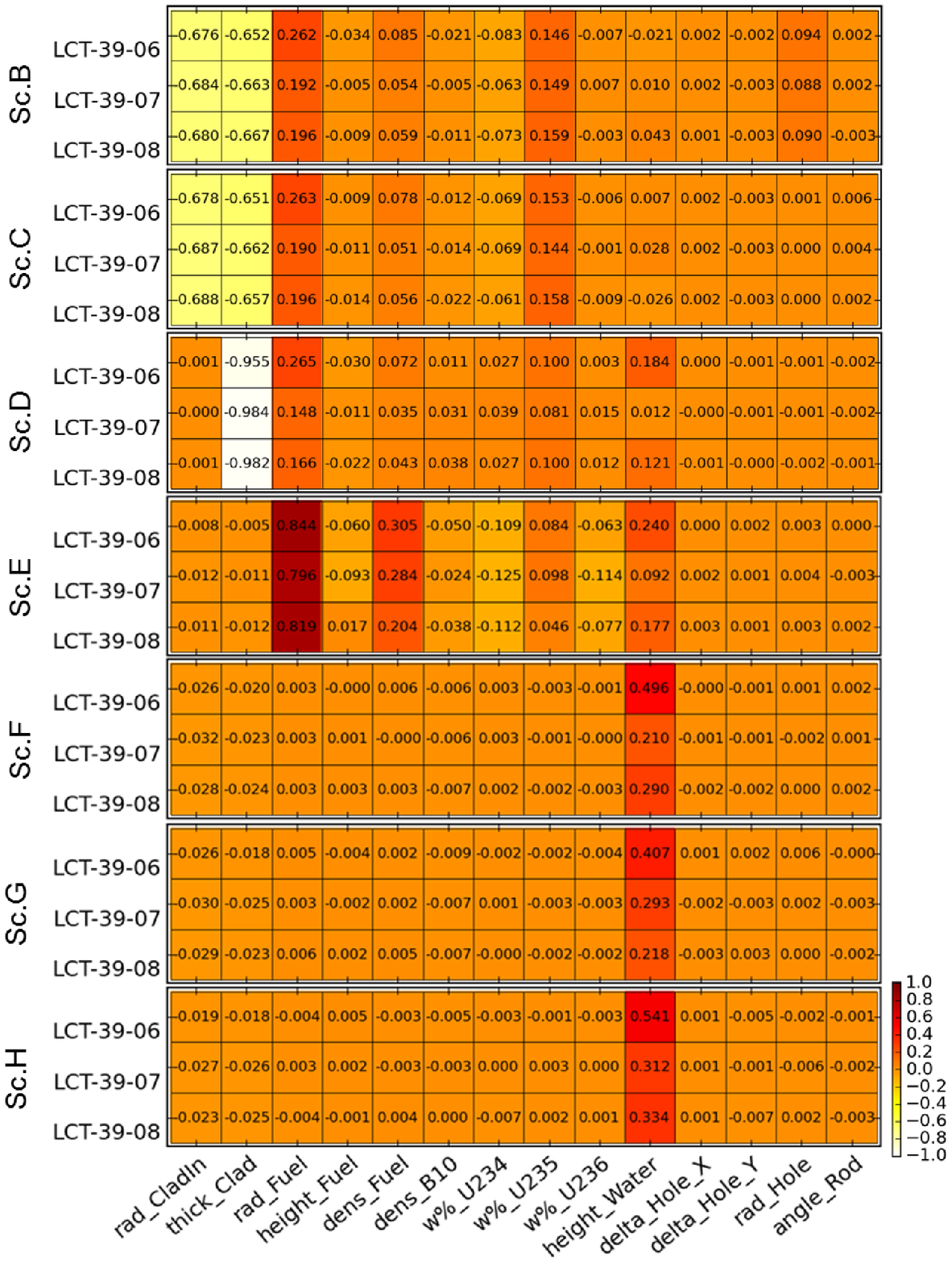}
	\caption{Correlation coefficients for each model parameter and the resulting 250 $k_{\rm eff}$ values. From top to bottom the scenarios B to H are shown. The increased number of parameters compared to the scenarios NoCor and A is due to the variations of single fuel rods. The color coded representation of the ${\rm cor}$ shows values from -1 to 1.}
	\label{fig:Fig8}
\end{figure}

%
%
\section{Discussion and Conclusions}
\label{Conclusions}

In this work we discussed nine different modeling approaches for a given set of experimental data, leading to different correlation coefficients and sensitivity profiles. Using a Monte Carlo approach, we calculated 250 samples for each experimental setup and scenario to obtain the resulting $k_{\rm eff}$ values (figure \ref{fig:keffs_StdDev}). The steering and analyzing of the SCALE6.1.2 CSAS5 sequences where done using SUnCISTT. 

Within each scenario we calculated for each pair of experiments the corresponding correlation coefficient and the 95$\%$ confidence intervals of this coefficient (figures \ref{fig:All_9_Sc_CI}, \ref{fig:ScVsCk}). We showed for each experiment and scenario the impact of the variation of each input parameter on the resulting $k_{\rm eff}$ by calculating the corresponding correlation coefficients (figure \ref{fig:Fig7}, \ref{fig:Fig8}).  

For the combination of water moderated, low enriched Uranium rods modeled with the criticality code KENO V-a we found that the correlation coefficients between the $k_{\rm eff}$'s of the experiments LCT-39 6, 7, 8 varied between 0 and 1 within the 95$\%$ confidence interval. 

The modeling assumption leading to scenario A and NoCor are identical, except that NoCor neglects correlations completely. The same holds for scenarios G and H, which are identical, but H neglects the correlations between experiments due to fuel similarities.

Varying all geometrical parameters affecting the outer cladding radius for each fuel rod separately leads to a significant decrease of the resulting correlation coefficient compared to the results derived from the assumption of all fuel rods having identical geometrical parameters (however, scenarios B, C and D still comprise high correlations). We found a significant drop of the ${\rm cor}$ value from scenario D to E (\ref{fig:ScVsCk}) as well as a significant drop of the $k_{\rm eff}$ uncertainty (figure \ref{fig:keffs_StdDev}). The main contribution to the $k_{\rm eff}$ uncertainty in scenario D stems from the cladding thickness, which in scenario E plays no role since it is varied for each fuel rod individually (figure \ref{fig:Fig8}).   

Scenarios F and G show a further drop of the correlation coefficient, but within the 95$\%$ confidence interval the ${\rm cor}$ values of the two scenarios overlap (figure \ref{fig:ScVsCk}). The difference of the two assumptions, knowing the exact position of each fuel rod for Scenario F or randomize their position in the grid for scenario G has a comparable smaller effect on ${\rm cor}$ than the assumption of a finite number of fuel rods. It is notable, that the sensitivity analysis shows the sole dependence of the $k_{\rm eff}$ uncertainty on the critical water height (figure \ref{fig:Fig8}). 

The different modeling assumptions might all be justified based on expert judgment. However, the sensitivity analysis reveals different sensitivity profiles, especially from scenario C to F. One could be tempted to choose the modeling assumptions based on the quality of the experimental data. As an example one could argue to choose scenario F or G, since the uncertainty of $k_{\rm eff}$ is much lower and the almost sole dependence of the $k_{\rm eff}$ uncertainty is on the critical water height. Following this argumentation, one could construct modeling assumptions based on the given experimental data to reduce uncertainties and to circumvent possible gaps in the data. But one has to be very careful with these options, and give very good arguments, why one chooses one scenario over another. The resulting covariance matrices directly influence the bias and its uncertainty, and thus the resulting upper sub-critical limit \cite{ Hoefer2015,ICNCBias, Sobes}.   

Using the covariance or correlation matrices for the purpose of validation or the determination of the upper sub-critical limit of an application case, the results can vary strongly, depending on the scenario. Following the argumentation of \cite{ICNCBias, Sobes}, a rule of thumb is that the higher the correlation coefficient, the lesser information is available, and thus the upper sub-critical limit decreases. This means, that being not able to distinguish between the different scenarios and identify the correct one based on the available data, one would in this case take the results associated with the highest correlation coefficient to get a more conservative estimate of the bias in code validation or the upper sub-critical limit.
   
Note that the underlying data for the work presented is partly constructed and fictive as it is a part of a calculation benchmark exercise \cite{UACSABMIV}. From the given data, any modeling assumption from scenario A to H could be justified. For further determination of the scenarios one would need to know e.g. if the fuel content and geometric description for each fuel rod was identical or if it varied. The statements presented above thus are only valid for the combinations of code and experiments discussed here.
 
To derive more general statements, further investigations have to be carried out. On the other hand, it may be problem dependent if and to what extent the regard for correlations between benchmark experiments could influence the bias determination. A sufficient number of statistical independent data sets, e.g. for experiments conducted in different laboratories using different materials, can always circumvent the problem of the correct determination of integral experimental covariance data. However, the accurate consideration of correlated data seems to be inevitable if the experimental data in a validation procedure is limited.
 
But even if one can avoid the determination of the accurate integral experimental covariance data due to statistically independent data sets, the selected modeling scenario should always be justified. The modeling assumptions have the potential to decrease the uncertainty of the resulting $k_{\rm eff}$ significantly.

\section{Acknowledgement}
Our work was financed by the German Federal Office for Radiation Protection under grant No.3614R03331. We thank W. J. Marshall and B. T. Rearden for the various fruitful discussions of our results and findings. We further would like to thank D. Mennerdahl for the discussions concerning the different scenarios.


\section{References}
\label{Ref}



\end{document}